**Nonlinear energy transfer in classical and quantum systems**

Leonid Manevitch[1], Agnessa Kovaleva[2]

[1] *Institute of Chemical Physics, Russian Academy of Sciences, Moscow 119991 Russia*
[2] *Space Research Institute, Russian Academy of Sciences, Moscow 117997 Russia*

We investigate the effect of slowly-varying parameter on the energy transfer in a system of weakly coupled nonlinear oscillators, with special attention to a mathematical analogy between the classical energy transfer and quantum transitions. For definiteness, we consider a system of two weakly coupled oscillators with cubic nonlinearity, in which the oscillator with constant parameters is excited by an initial impulse, while a coupled oscillator with slowly-varying parameters is initially at rest. It is proved that the equations of the slow passage through resonance in this system are identical to equations of the nonlinear Landau-Zener (LZ) tunneling. Three types of dynamical behavior are distinguished, namely, quasi-linear, moderately nonlinear and strongly nonlinear. The quasi-linear systems exhibit a gradual energy transfer from the excited to the attached oscillator, while the moderately nonlinear systems are characterized by an abrupt transition from the energy localization on the excited oscillator to the localization on the attached oscillator. In the strongly nonlinear systems, the transition from the energy localization to strong energy exchange between the oscillators is revealed. A special case of the rapid irreversible energy transfer in the strongly nonlinear system with slowly-varying parameters is also investigated. The conditions providing different types of the dynamical behavior are derived. Explicit approximate solutions describing the transient processes in moderately and strongly nonlinear systems are suggested. Correctness of the constructed approximations is confirmed by numerical results.

## 1. Introduction

In the work, we develop an analytical framework to understand the effect of the slow change of frequencies on targeted energy transfer (TET) in a system of two weakly-coupled nonlinear oscillators, with special attention to an analogy between the energy transfer in a classical oscillatory system and nonlinear Landau-Zener (LZ) two-level tunneling.

The concept of TET in nonlinear classical systems with constant coefficients was introduced earlier [1, 2]; motivated by this concept, a common consideration of classical and quantum energy transfers was suggested [3]. Classical (or quantum) TET have interesting and nontrivial applications in diverse fields of applied mathematics, natural sciences, and engineering, and a large number of examples and the discussion of recent advances in the theory and applications of the TET phenomenon have been reported in the literature, e.g., in [4–15] and references therein. Most of the theoretical results concern the systems with constant parameters but over last decade nonlinear quantum transition under slow driving has been investigated using different methods [11 – 21].

It was recently shown that the equations of the slow dynamics in passage through resonance in the linear [8–10] and nonlinear [11] systems of two weakly coupled oscillators with slowly-varying frequencies are equivalent to the well-known Landau-Zener (LZ) equations [22, 23]. However, even in the linear case an exact solution to the Landau-Zener equation is too complicated for a straightforward analysis. It is not surprising, then, that the study of nonlinear tunneling was mainly focused on the calculation of the jump in the adiabatic invariant at an instant of tunneling but the transient processes before and after tunneling still remain uninspected. The purpose of the current paper is to develop a procedure for an adequate analytical description of the transient processes in both classical and quantum nonlinear systems with slowly time-dependent parameters.

The paper is organized as follows. In the first part of the paper (Sections II and III), the quasi-resonant dynamics of a system of two weakly coupled nonlinear oscillators with *constant* parameters is studied. In Sec. II, the appropriately defined slow dynamics for different sets of parameters are investigated. Note that the systems similar to that one considered in Sec. II were investigated and the phase space was presented, e.g., in [16, 17, 24, 25]. However, a more detailed analysis of the dynamical transitions given in Sec. II provides a deeper understanding of the underlying dynamics both for the conservative system as well as for nonlinear time-dependent tunneling. In particular, the examination of the dynamical portraits (Sec. II) reveals three types of the dynamical behavior, quasi-linear, moderately nonlinear, and strongly nonlinear ones, while the previous works on nonlinear tunneling have drawn attention to a single parametric boundary, associated with the change of the number of the stationary points and corresponding, in our terminology, to the boundary between the quasi-linear and moderately nonlinear behavior (cf. [16, 17]). Furthermore, as shown in Sections II and III, the dynamical transition turns out to be closely related to the behavior of the so-called Limiting Phase Trajectory (LPT) [26]. This implies that the analysis of the dynamical behavior cannot be reduced to the commonly accepted study of the stationary points and their evolution.

A difference between the ensembles of trajectories corresponding to different types of motion is discussed in Sec. II; the relationship between the parameters determining the boundary between the moderately and strongly nonlinear systems is derived in Sec III.

In the rest of the paper, the transient processes in the system with slowly-varying coefficients are examined. Section IV suggests an explicit asymptotic solution describing the transient processes in the moderately nonlinear system under adiabatically slow driving. The approach is based on the fact that at each stage of motion, both before and after tunneling, the energy of the system is localized on one of the oscillators, while the energy of the second one is small enough. This inherent property allows using an iterative procedure earlier developed for quasi-linear systems [11]. The initial iteration expressed through the Fresnel integral formally explains the convergence of energy to a certain limiting value, while the successive iterations improve the accuracy of the approximate solution. In addition to the description of the transient processes, the explicit approximation allows the analytical calculation of the change of energy in tunneling. The numerical results show that even the main approximation provides an adequate representation of the transient processes.

Section V presents an approximate analysis of energy transfer in the strongly nonlinear system under slow driving. Details of this analysis are provided in Appendix. A special case of the rapid irreversible energy transfer in the strongly nonlinear system with slowly-varying parameters is also investigated. Concluding remarks are given in Sec. VI.

## II. STATIONARY MODEL

In this part of the paper, the dynamics of two weakly coupled nonlinear oscillators with *constant* parameters is studied in details. The results of this section provide some background information that will be applied in the further analysis of nonlinear tunneling. The task is to highlight the distinguishing dynamical properties of the system and find the sets of the parameters corresponding to quasi-linear, moderately nonlinear and strongly nonlinear types of the dynamical behavior. The obtained results open avenues for explaining the dynamical transitions in the systems with adiabatically changed parameters.

The system under consideration consists of two oscillators with cubic restoring forces connected with linear coupling. The first oscillator is excited by an initial impulse, while a coupled oscillator is initially at the rest. We denote by $u_1$ and $u_2$ the absolute displacements of the 1st and 2nd oscillators, respectively; by $m$, the equal masses of the oscillators; by $c_1$ and $c_2$, the coefficients of linear stiffness of the corresponding oscillators; by $c_3$, the coefficient of cubic nonlinearity; by $c_{12}$, stiffness of linear coupling. The nonlinear equations of motion are given by

$$m\frac{d^2u_1}{dt^2} + c_1u_1 + c_3u_1^3 + c_{12}(u_1 - u_2) = 0, \tag{2.1}$$

$$m\frac{d^2u_2}{dt^2} + c_2u_2 + c_3u_1^3 + c_{12}(u_2 - u_1) = 0,$$

The small parameter of the system $\varepsilon$ is defined as the relative stiffness of weak coupling: $c_{12}/c_1 = 2\varepsilon << 1$. Assuming weak nonlinearity and taking into account resonance properties of the system, the system parameters are redefined as follows:

$$c_1/m = \omega_0^2,\ \tau_0 = \omega_0 t,\ c_2/m = \omega_0^2(1 + 2\varepsilon g),\ c_3/c_1 = 8\varepsilon\alpha,\ c_{12}/c_r = 2\varepsilon\lambda_r,\ r = 1, 2, \tag{2.2}$$

where $g$ = const. It follows from definition (2.2) that $\lambda_1 = 1$, $\lambda_2 = (1 + 2\varepsilon g)^{-1}$. Substituting equalities (2.2) into Eq. (2.1) and ignoring the terms of order higher than $\varepsilon$, we obtain the rescaled dimensionless system

$$\frac{d^2u_1}{d\tau_0^2} + u_1 + 2\varepsilon(u_1 - u_2) + 8\varepsilon\alpha u_1^3 = 0, \tag{2.3}$$

$$\frac{d^2u_2}{d\tau_0^2} + (1 + 2\varepsilon g)u_2 + 2\varepsilon(u_2 - u_1) + 8\varepsilon\alpha u_2^3 = 0.$$

System (2.3) is assumed to be initially at rest, with a unit impulse applied to the first oscillator. The corresponding initial conditions are given by

$$\tau_0 = 0,\ u_1 = u_2 = 0;\ v_1 = du_1/d\tau_0 = 1,\ v_2 = du_2/d\tau_0 = 0.$$

If a unit impulse is applied to the coupled oscillator but the first one is initially at rest, then the initial conditions are written as

$$\tau_0 = 0,\ u_1 = u_2 = 0;\ v_1 = du_1/d\tau_0 = 0,\ v_2 = du_2/d\tau_0 = 1.$$

The chosen initial conditions determine the limiting phase trajectory (LPT). The LPT concept associated with the complete energy exchange between the oscillators in the symmetric system ($g$ = 0) was introduced in [26]. A more detailed analysis of the LPT properties in the asymmetric system ($g \neq 0$) will be given below.

An asymptotic solution of system (2.3) for small $\varepsilon$ is constructed in the form of the multiple-scale expansion. We introduce the complex-valued change of variables

$$f_r = (v_r + iu_r)\,e^{-i\tau_0}\,,\; f_r^* = (v_r - iu_r)\,e^{i\tau_0}\,,\; r = 1, 2. \tag{2.4}$$

where the functions $f_r$ and their time-derivative are sought as

$$f_r(\tau_0,\ \tau_1,\ \ldots) = f_{r0}(\tau_0,\ \tau_1,\ \ldots) + \varepsilon f_{r1}(\tau_0,\ \tau_1,\ \ldots) + O(\varepsilon^2), \tag{2.5}$$

$$\frac{df_r}{d\tau_0} = \frac{\partial f_r}{\partial \tau_0} + \varepsilon\,\frac{\partial f_r}{\partial \tau_1} + O(\varepsilon^2),\ \tau_1 = \varepsilon\tau_0.$$

Reproducing the transformations of [26] we find that

$$f_{10}(\tau_1) = a(\tau_1)\,e^{i\tau_1}\,,\; f_{20}(\tau_1) = b(\tau_1)\,e^{i\tau_1}\,, \tag{2.6}$$

where the envelopes $a(\tau_1)$ and $b(\tau_1)$ satisfy the equations

$$\frac{da}{d\tau_1} + ib - 3i\alpha|a|^2 a = 0, \tag{2.7}$$

$$\frac{db}{d\tau_1} + ia - 3i\alpha|b|^2 b - 2igb = 0,$$

$$a(0) = 1,\ b(0) = 0.$$

It easily follows from Eqs. (2.4) - (2.7) that energy of the first and second oscillators can be approximated as $e_1 = ½|a|^2$, $e_2 = ½|b|^2$; furthermore, $|a|^2 + |b|^2 = 1$, $e_1 + e_2 = 1/2$. We recall that Eqs. (2.7) are identical to the equations of two-state atomic tunneling [24], thereby confirming a direct mathematical analogy between quantum and classical tunneling.

The change of variables

$$a = \cos\theta e^{i\delta_1}\,,\; b = \sin\theta e^{i\delta_2}\,,\; \Delta = \delta_1 - \delta_2, \tag{2.8}$$

transforms system (2.7) to the real-valued equations

$$\frac{d\theta}{d\tau_1} = \sin\Delta, \tag{2.9}$$

$$\sin 2\theta \frac{d\Delta}{d\tau_1} = 2(\cos\Delta + 2k\sin 2\theta)\cos 2\theta - 2g\sin 2\theta,$$

where $k = 3\alpha/4$. In a simple mechanical analogy, system (2.9) describes motion of an "equivalent" non-rigid pendulum with tilt angle $\theta$, see e.g. [24, 25]. The initial conditions for system (2.9) are given either as $\theta = 0$, $\Delta = \pi/2$ (the first oscillator is excited but the second one is initially at rest) or as $\theta = \pi/2$, $\Delta = \pi/2$ (the second oscillator is excited but the first one is at rest). Note that both conditions correspond to the LPTs of system (2.9).

We first define the steady states of system (2.9). The stationarity conditions $d\theta/d\tau_1 = 0$, $d\Delta/d\tau_1 = 0$, gives $\sin\Delta = 0$ Therefore, all steady states of lie on the vertical axes $\Delta_1 = 0$ and $\Delta_2 = \pi$; the corresponding values of $\theta$ are given by

$$(\pm 1 + 2k\sin 2\theta)\cot 2\theta = g, \tag{2.10}$$

where the signs "+" and " –" correspond to $\Delta = 0$ and $\Delta = \pi$, respectively. If $k < 0.5$, then there exists a unique solution of Eq. (2.10) on each of the axes $\Delta = 0$ and $\Delta = \pi$. It was recently shown [11] that the phase portrait of system (2.9) with $0 < k < 0.5$ remains similar to the phase portrait of its linear counterpart; furthermore, if the condition $0 < k < 0.5$ holds, then the solution of the nonlinear system is close to that of the linear system. The proximity of the solutions allows us to consider the systems with the parameter of nonlinearity $0 < k < 0.5$ as *quasi-linear*. Below we focus on the system with the parameter $k > 0.5$.

We recall that the systems similar to (2.7) or (2.9) were investigated and the phase space was presented, e.g., in [16, 17, 24, 25]. The only critical boundary reported in the literature [16, 17] corresponds to $k = 0.5$; it is associated with the change of the number of the stationary states on the axis $\Delta = \pi$. Here we analyze the dynamical transitions in more detail with the aim of highlighting different types of the dynamical behavior depending on both parameters $k$ and $g$.

As seen in Figs. 1 – 4, if $k > 0.5$, then there exists a certain value $g^*$ such that the system possesses 3 stationary points on the axis $\Delta = \pi$ if $|g| < g^*$ but it has a single point if $|g| > g^*$. Referring to Eq. (2.10), we denote $F(\theta) = (-1 + 2k\sin 2\theta)\mathrm{ctg}2\theta$, $\theta_T = \arg\{\max F(\theta)\}$, $\theta \in [0, \pi/2]$. It is obvious that two stationary points coalesce if $F(\theta_T) = g^*$. The maximum condition $dF/d\theta = 0$ at $\theta = \theta_T$ shows that the point of coalescence of two stationary states $\theta_T$ and the critical parameter $g^*$ are given by

$$\sin 2\theta_T = (2k)^{-1/3}, \; g^* = \pm [(2k)^{2/3} - 1]^{3/2}. \quad (2.11)$$

Note that $g^*$ coincides with the critical parameter $\gamma_c$ derived in [17] (in other notations).

Phase portraits of system (2.9) with $k > 0.5$ are given in Figs. 1 – 4. It is seen that, in addition to a single stable center on the axis $\Delta = 0$, on the axis $\Delta = \pi$ the system has two stable centers separated a hyperbolic point. We demonstrate that the system with 4 stationary states exhibits two different types of dynamical behavior, depending on the behavior of the separatrix passing through the hyperbolic point. In the current section we demonstrate that the system with four stationary states exhibits two different types of dynamical behavior, depending on the behavior of the separatrix passing through the hyperbolic point. It is necessary to note that, besides the stationary points and their evolution, a special attention should be given to the behavior of the LPTs.

As a first example, we consider the system with $k = 0.65$. It is easy to deduce from Eqs. (2.9) that the change of the sign of the parameter $g \to -g$ entails the change of the solution such that $\theta \to \pi - \theta$, $\Delta \to 2\pi - \Delta$. This allows us to construct the phase portraits only for $g \ge 0$ (Fig.2). Bold lines in Fig. 1 depict the LPT of system (2.9).

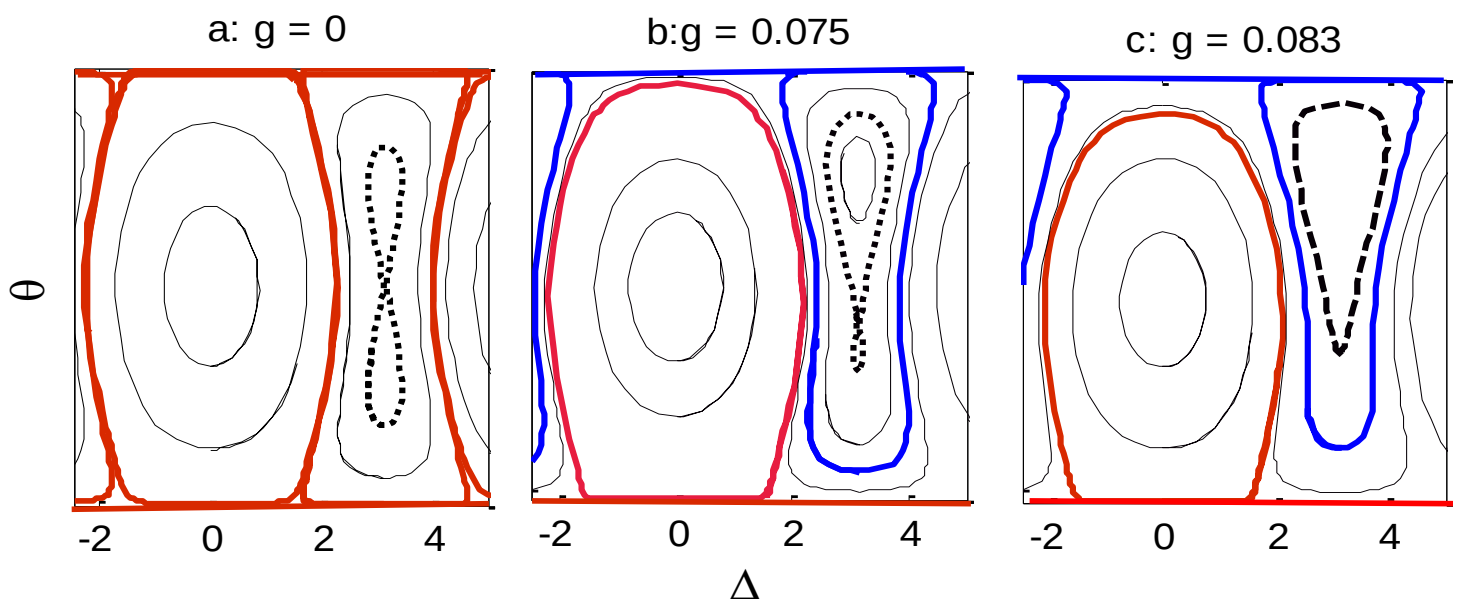

FIG. 1. Phase portraits of system (2.9) with $k = 0.65$; with an increase of $g$, the homoclinic separatrix (dotted lines in Fig. 1$a$ and Fig. 1$b$) is transformed into a closed orbit; dashed line (Fig. 1$c$) depicts the orbits emerging at annihilation of the lower stable center and unstable hyperbolic point

Figure 1*a* illustrates complete energy exchange between the symmetric oscillators ($g$ = 0), i.e., the upper level $\theta = \pi/2$ is reached during the cycle of motion along the LPT starting at $\theta = 0$. Motion along the closed orbits within the domain encircled by the LPT obviously provides less extensive energy exchange than motion along the LPT.

Dotted lines in Fig. 1 correspond to the homoclinic separatrix. It is seen that with an increase of $g$ the lower homoclinic loop vanishes through the coalescence of the stable and unstable states, and the number of the stationary states changes from 4 to 2. The corresponding critical value $g^* = 0.083$ coincides with the theoretical value given by formula (2.11). Similarly, one can conclude that the number of the stationary points changes from 2 to 4 at $g^* = -0.083$. Vanishing of the homoclinic separatrix with the occurrence of a new closed orbit of finite period (Fig. 1*c*) characterizes *moderately nonlinear* systems, in which the energy localization near the lower center changes to the localization near the upper center

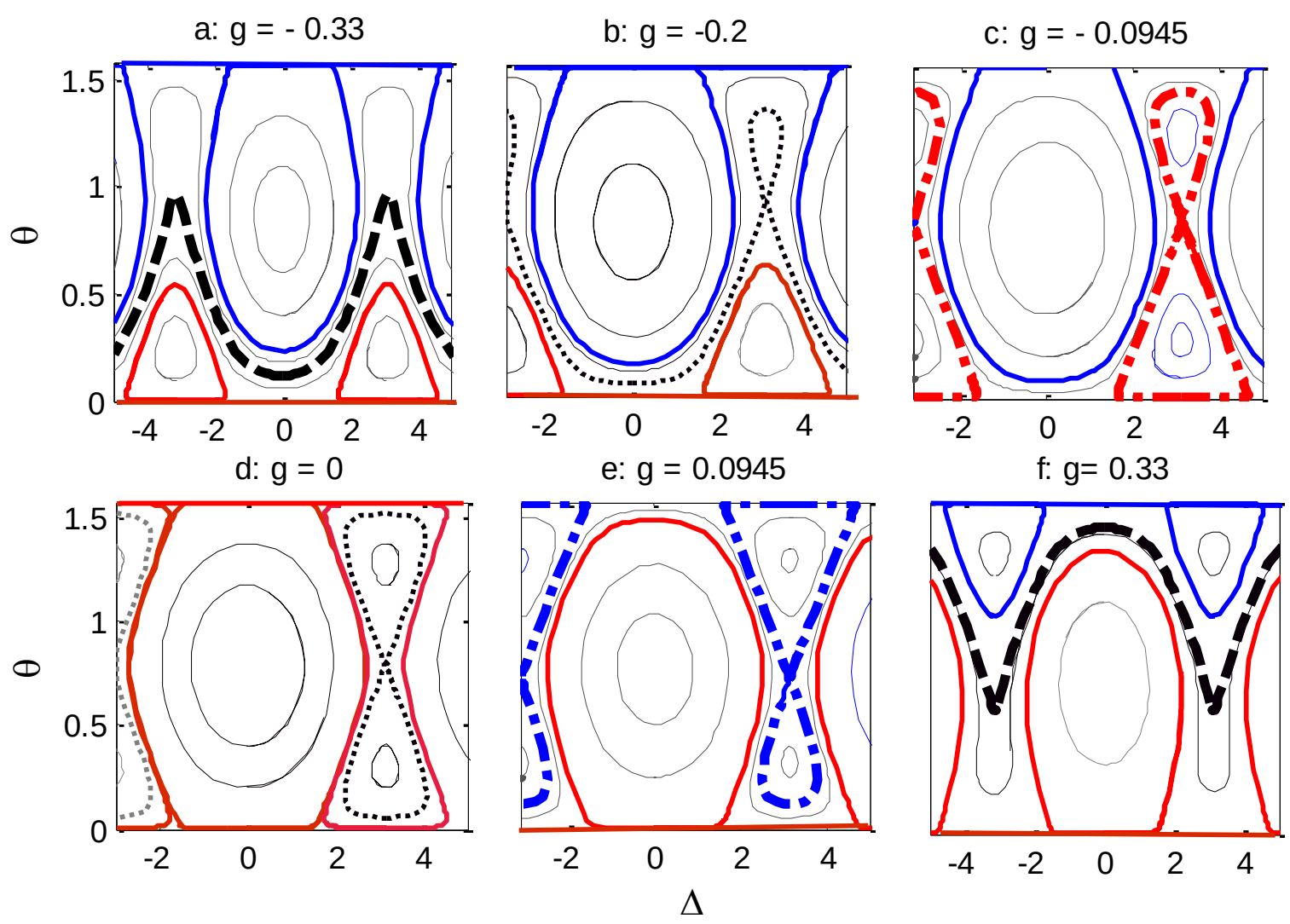

FIG. 2. Phase portraits of system (2.9) with $k = 0.9$: bold lines correspond to LPTs; dash-dotted lines depict the homoclinic separatrix coinciding with the LPT; edges of the dashed “beaks” rest at the points of annihilation of the stable and unstable states

As seen in Fig. 2, a system with $k = 0.9$ exhibits a more complicated dynamical behavior. For clarity, the phase portraits for both $g < 0$ and $g > 0$ are shown. The change from 2 to 4 fixed points at $g = -0.33$ (Fig. 2*a*) leads to the emergence of a separatrix passing through the hyperbolic point and consisting of the homoclinic and heteroclinic branches (Fig. 2*b*); at $g = -0.0945$, the heteroclinic loop coincides with the LPT at $\theta = 0$ (Fig. 2*c*). Further increase of $g$ results in the appearance of the homoclinic separatrix (Fig.

2*d*) and then to the confluence of the separatrix with the LPT at $\theta = \pi/2$ for $g = 0.0945$ (Fig. 2*e*). At $g > 0.0945$, the upper branch of the LPT turns into a heteroclinic separatrix similar to that in Fig. 2*b*; finally, the degeneration of the separatrix with the change from 4 to 2 steady states is observed (Fig. 2*f*). The numerical value $|g^*| = 0.33$ coincides with the results of calculation by formula (2.11). Figure 2*d* also illustrates complete energy exchange in the symmetric system moving along the LPT.

Figure 3 presents the phase portraits of the system with $k = 1$, $g > 0$. It is seen in Fig. 3*a* that for $g = 0$ the LPT (bold dash-dotted line) includes the coinciding segments of the homoclinic and heteroclinic separatrices; further increase of $g$ leads to the transformation of a homoclinic loop into a heteroclinic loop and then into an unlocked orbit. The phase portraits for $g < 0$ can be constructed by symmetry.

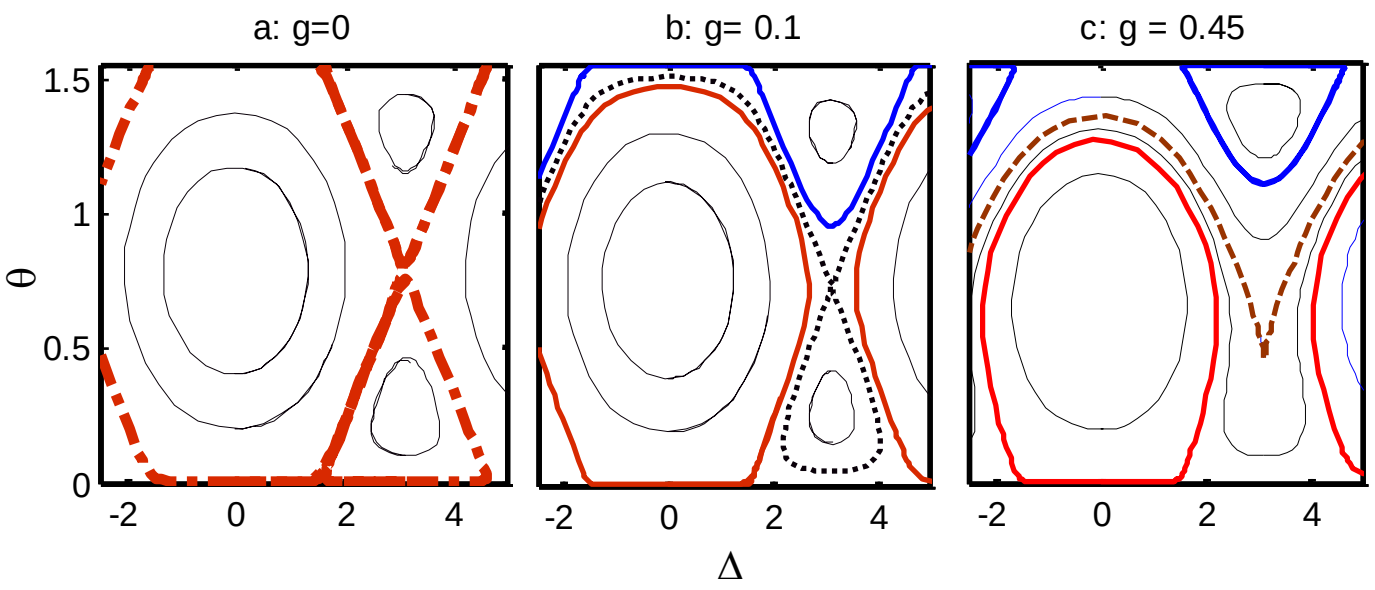

FIG. 3. Phase portraits of system (2.9) with $k = 1$; dash-dotted in Fig. 3*a* depicts the separatrix coinciding with the LPT. An increase of $g$ leads to the transformation of the upper homoclinic loop into the heteroclinic loop (dotted line in Fig. 3*b*) and then into an unlocked orbit (dashed line in Fig. 3*c*)

Figure 4 demonstrates the transformation of the heteroclinic separatrix into the LPT in the case of $k = 1.1$. In both cases $k = 1$ and $k > 1$, the transition from energy localization near the stable state to energy exchange occurs due to the coalescence of the stable and unstable states (Figs. 3, 4) resulting in the formation of an unlocked orbit.

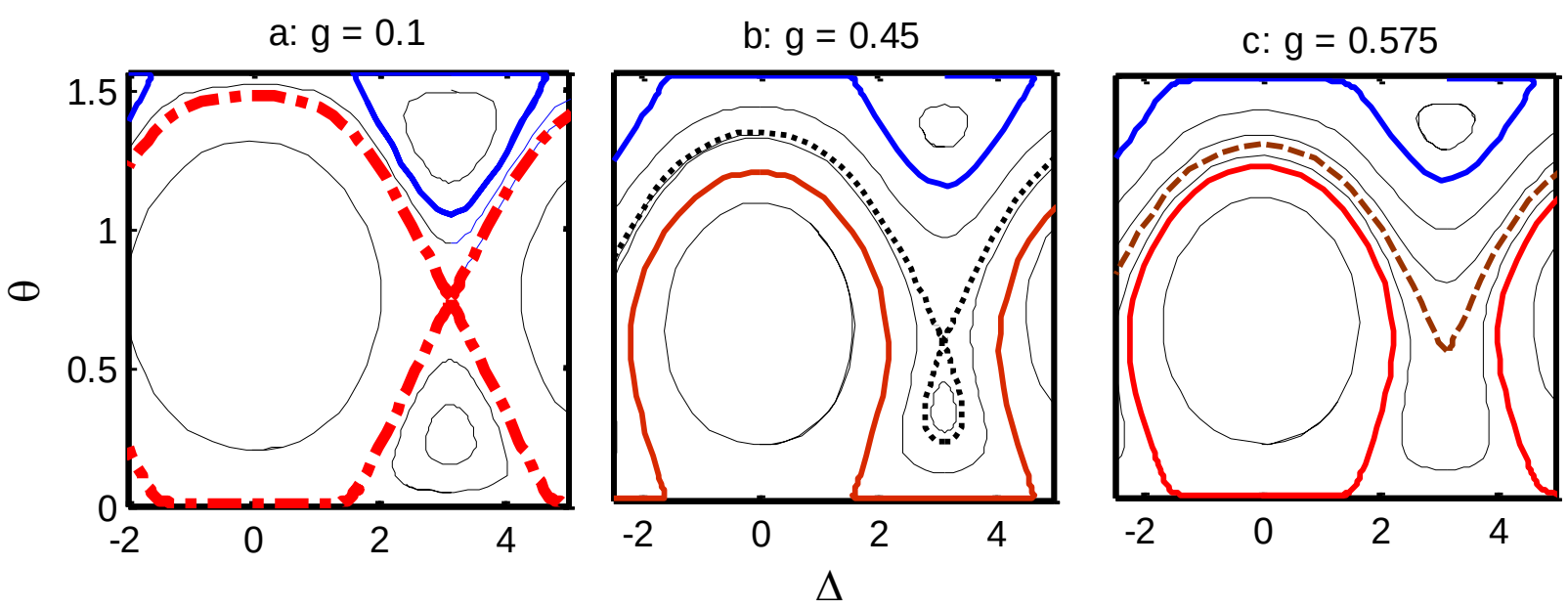

FIG. 4. Phase portraits of system (2.9) with $k = 1.1$: dotted lines depict the separatrices; bold dash-dotted line in Fig. 4*a* depicts the heteroclinic separatrix coinciding with the LPT at $g = 0.1$. An increase of $g$ leads to the formation of a new separatrix (Fig. 1*b*) with further transformation into an unlocked orbit (Fig. 1*c*)

The system featuring the coincidence of the separatrix with the LPT and the following transitions is referred to as *strongly nonlinear systems.* The coalescence of the stable and unstable states gives rise to a new unlocked orbit, motion along which can be interpreted as rotations of the “equivalent pendulum” (2.9) in the outer domain near the LPT (Fig. 2– 4). It will be shown in Sec. III that the system may exhibit strongly nonlinear behavior if $k > 0.77$.

## III. CRITICAL PARAMETERS

In this section we find the critical relationships between the parameters $k$ and $g$ that determines a boundary between the moderately nonlinear and strongly nonlinear dynamical behavior. As remarked previously, the system may be considered as strongly nonlinear if its separatrix coincides with the LPT. Now we find prerequisites for the existence of such a trajectory.

It is easy to check that system (2.9) preserves the integral of motion

$$K = (\cos\Delta + k\sin 2\theta)\sin 2\theta + g\cos 2\theta. \tag{3.1}$$

The initial condition $\theta = 0$ yield $K = g$ and, therefore,

$$(\cos\Delta + k\sin 2\theta)\sin 2\theta - 2g\sin^2\theta = 0 \tag{3.2}$$

on the LPT. It now follows from Eqs. (2.9) and (3.2) that

$$d\theta/d\tau_1 = V = \sin\Delta\,;\;\; V = \pm\,[1 - (k\sin 2\theta - g\tan\theta)^2]^{1/2}. \tag{3.3}$$

Using Eq. (3.2) to exclude $\Delta$, we replace system (2.9) by the following second-order equation:

$$\frac{d^2\theta}{d\tau_1^2} + \frac{dU}{d\theta} = 0 \tag{3.4}$$

with $\theta(0) = 0$, $V(0) = 1$. The potential $U(\theta)$ in (3.4) can be found from the energy conservation law $E = ½V^2 + U(\theta) = 1$. We thus obtain

$$U(\theta) = 1 - ½V^2 = ½[1 + (k\sin 2\theta - g\tan\theta)^2], \tag{3.5}$$

and, therefore, the maximum value $U(\theta) = 1$ is attained at $V = 0$.

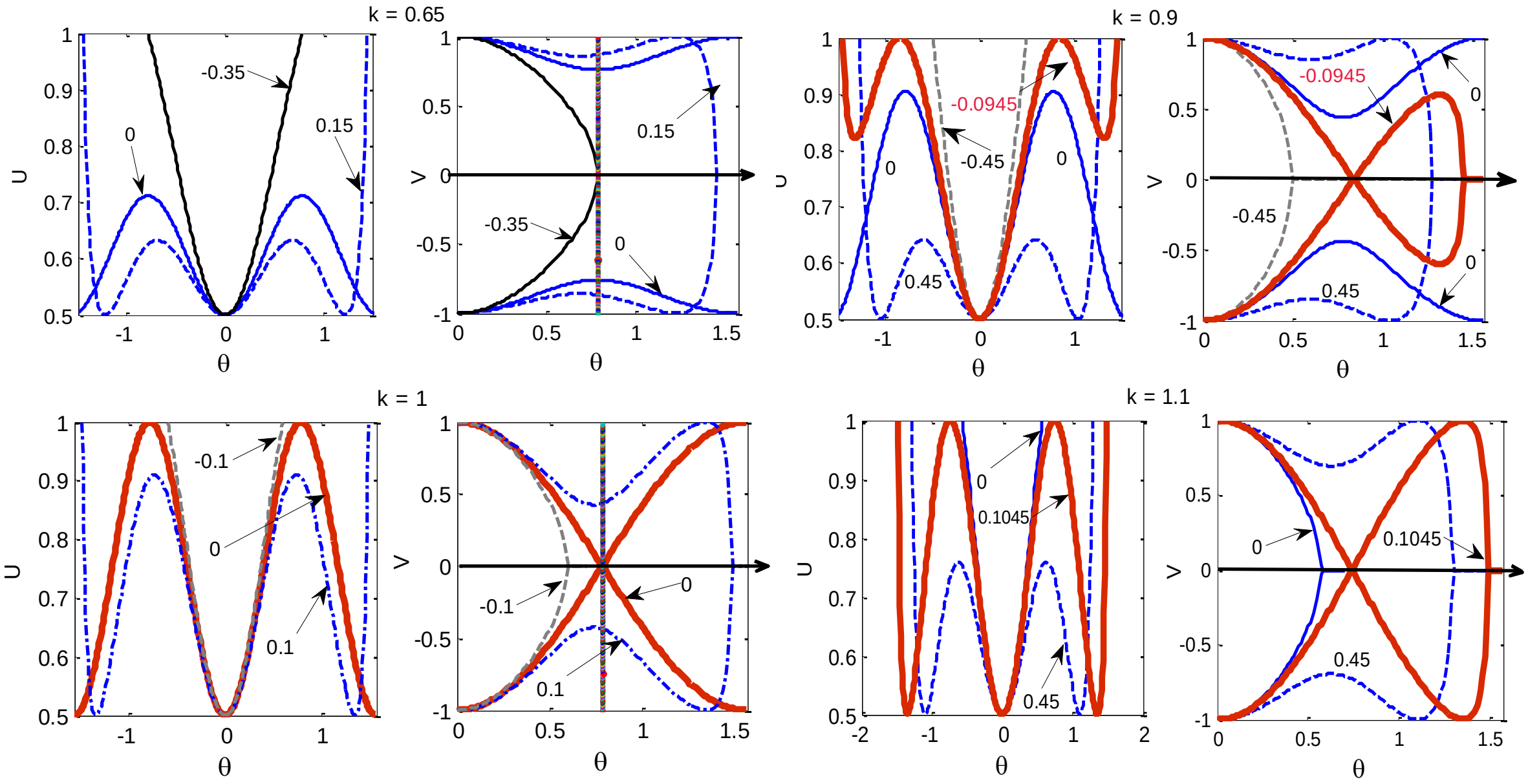

FIG. 5. Potential $U(\theta)$ and phase planes for system (3.2) with different coefficients of nonlinearity; detuning is indicated on each curve; bold lines depict the critical potentials and the corresponding separatrices confluent with the LPT at $\theta = 0$

The sought separatrix may exist if and only if $dU/d\theta = 0$ at $\theta_h \in (0, \pi/2)$, as in this case there exists a potential barrier corresponding to the local maximum $U(\theta_h)$ and attained at $V = 0$; the latter condition is equivalent to $\Delta = 0$. This implies that $(\theta_h, 0)$ is a hyperbolic point. It now follows from (3.5) that the equality $dU/d\theta^* = 0$ is equivalent to

$$(k\sin 2\theta_h - g\tan\theta_h)(2k\cos 2\theta_h - g/\cos^2\theta_h) = 0. \tag{3.6}$$

It follows from (3.3) that $(k\sin 2\theta_h - g\tan\theta_h)^2 = 1$at $V = 0$. It now follows from Eq. (3.6) that

$$4k\cos^4\theta_h - 2k\cos^2\theta_h - g = 0,\ \cos^2\theta_h = \frac{1}{4} \pm \sqrt{\frac{1}{16} + \frac{g}{4k}}\,. \tag{3.7}$$

If $g/k << ¼$, then $\cos^2\theta_h \approx 1/2 + g/2k$, $\theta_h \approx \pi/4 - g/2k$. Using this approximation, we obtain from Eq. (3.8) a simple condition for the existence of the required separatrix:

$$|k - g_h| = 1. \tag{3.9}$$

It is easy to check that the theoretical threshold $g_h$ closely agrees with the results of numerical calculations. In particular, for $k = 0.9$ we have $g_h = -\ 0.1$, whereas the numerical threshold $g = -\ 0.0945$ (Fig. 2); for $k = 1.1$ we have $g = g_h = 0.1$ (Fig. 4).

If $g < 0$, then the solution (3.9) exists provided $|g| \le k/4$; in the limiting case $g = -k/4$ we have $\cos\theta_h = ½$, $\theta_h = \pi/3$, that is condition (3.6) becomes

$$\frac{3\sqrt{3}}{4}k = 1,\ k^* \approx 0.77. \tag{3.10}$$

The inequalities $k \ge k^*$, $|g| \le k/4$ express *the necessary condition* for the existence of the separatrix coinciding with the LPT at $\theta = 0$. Additionally, one needs to calculate the argument $\theta_h$ by Eq. (3.9) and check condition (3.6) at $\theta = \theta_h$. In the same way, one can prove that if the initial condition is taken at $\theta = \pi/2$, then the condition (3.9) is turned into the equality

$$|k + g_h| = 1. \tag{3.10}$$

## IV. TUNNELING IN MODERATELY NONLINEAR SYSTEMS

As mentioned in the introductory section, an effect of time-dependent parameters on energy transfer in classical systems was examined only for linear and quasilinear oscillators [8 − 11]. In this section, we analyze energy transfer in a system of two *moderately nonlinear* weakly coupled oscillators with slowly-varying parameters. The main goals are to demonstrate an exact mathematical analogy between energy transfer in a classical system and nonlinear Landau-Zener tunneling and develop an asymptotic procedure for an adequate analytical description of transient processes in both classical and quantum systems.

It is well-known that energy transfer in a nonlinear system with adiabatically changed parameters is divided into two stages, with each characterizing by adiabatic invariance but separated by an abrupt jump at an instant of tunneling. Adiabaticity breaking under slow driving was intensively studied over last decades using various approximations [16–21]. However, an explicit description of the transient processes in non-autonomous nonlinear systems is prohibitively difficult, and the study of adiabatic tunneling was focused on the calculation of the jump of the adiabatic invariant at tunneling. This section is focused on the analytical investigation of transient processes before and after tunneling. The asymptotic analysis accounts for the fact that in the first interval of motion the most part of energy is localized on the excited oscillator but the residual energy of the coupled oscillator is small enough; after tunneling, localization of energy takes place on the coupled oscillator but energy of the initially excited oscillator becomes small. An introduction of a small parameter characterizing a relative energy level allows the construction of an explicit asymptotic solution. In the current section, we derive an explicit asymptotic representation of energy both before and after tunneling in terms of the Fresnel integrals; this solution formally explains the convergence of the trajectory to a certain limiting point. The approximate solution is used to estimate the change of energy due to tunneling.

As in Sec. II, we consider a system of two weakly-coupled oscillators. We assume here that linear stiffness of the coupled oscillator is a slow function of time, namely $c_2(\tau_2) = c_1(1 + 2\varepsilon g(\tau_2))$, where $g(\tau_2) = g_0 + g_1\tau_2$, $\tau_2 = \varepsilon\tau_1$. Reproducing the transformations (2.2) – (2.5), we reduce the equations of motion to the form similar to Eqs. (2.7) but involving the time-dependent detuning $g(\tau_2)$

$$\frac{da}{d\tau_1} + ib - 3i\alpha|a|^2 a = 0, \qquad (4.1)$$

$$\frac{db}{d\tau_1} + ia - 3i\alpha|b|^2 b + 2ig(\tau_2)b = 0,$$

$$a(0) = 1,\ b(0) = 0,$$

the coefficient of nonlinearity $\alpha$ is determined as in (2.2). We note that system (4.1) may be considered as a nonlinear analog of the well-known Landau-Zener equations of quantum tunneling with the initial conditions at $\tau_1 = 0$. This identity allows an extension of the approaches developed for the study of energy transfer in classical oscillatory systems to nonlinear quantum Landau-Zener tunneling.

As shown in Sec. II, the functions $|a|^2$ and $|b|^2$ can be considered as the main approximations to the energy of the first and second oscillators, respectively: $e_1 = ½|a|^2$, $e_2 = ½|b|^2$; furthermore, it is easy to check that the conservation law $|a|^2 + |b|^2 = 1$ holds true for system (4.1) despite its non-stationarity.

The change of variables $a = \cos\theta e^{i\delta_1}$, $b = \sin\theta e^{i\delta_2}$, $\Delta = \delta_1 - \delta_1$ reduces system (4.1) to the equations similar to (2.9)

$$\frac{d\theta}{d\tau_1} = \sin\Delta, \qquad (4.2)$$

$$\sin 2\theta\frac{d\Delta}{d\tau_1} = 2(\cos\Delta + 2k\sin 2\theta)\cos 2\theta - 2g(\tau_2)\sin 2\theta,$$

$$\theta(0) = 0,\ \Delta(0) = \pi/2.$$

Numerical results for systems (4.1) (4.2) with $k$ = 0.65, $g_0 = -0.5$, $\varepsilon g_1$ = 0.001 are given in Fig. 7. The phase portrait of system (4.2) and the plots of $|a|^2$ and $|b|^2$ clearly demonstrate the occurrence of adiabatic tunneling at $T \approx 585$. The corresponding value of critical detuning $g$ = 0.085 is close to the theoretical value $g^* \approx 0.083$. Note that the parameters of numerical simulations in this and next sections are chosen for illustrative purposes but the qualitative features of the results hold true for a wide range of parameters.

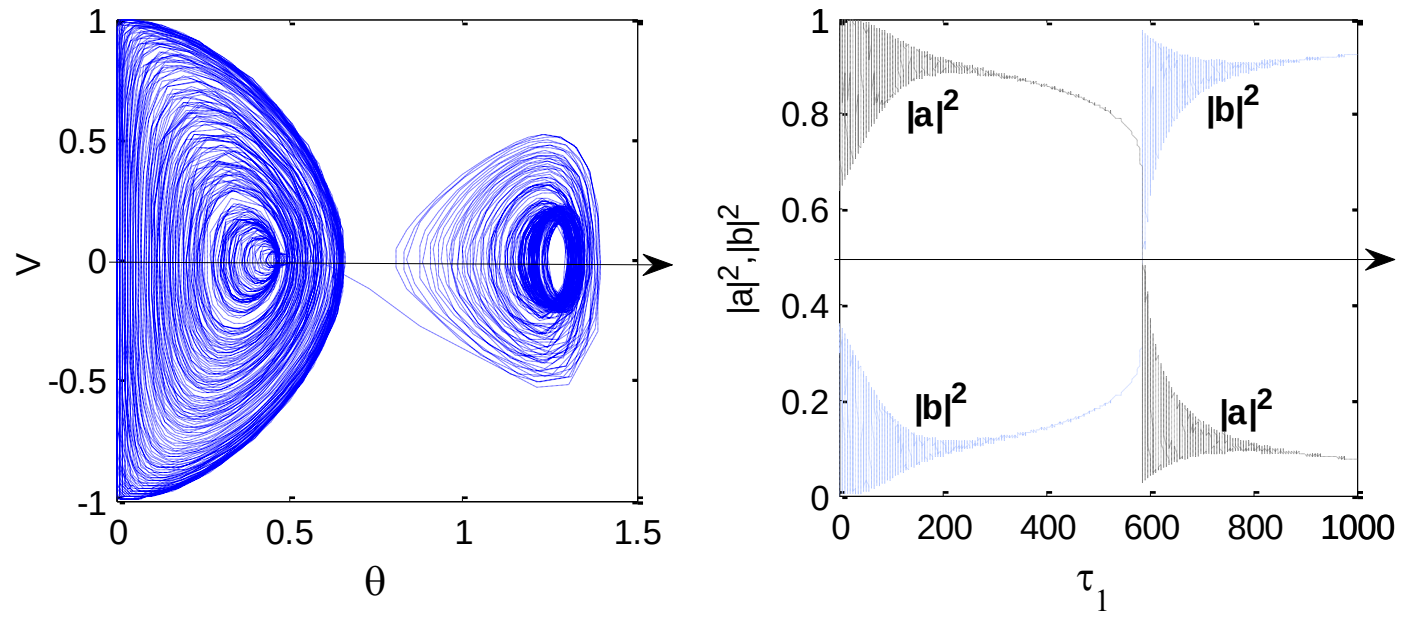

***a)*** ***b)***

FIG. 6. Dynamical behavior of system (4.2) on the interval $0 \le \tau_1 \le 1000$:
($a$): phase portrait in the plane ($\theta$, $V = d\theta/d\tau_1$); ($b$): plots of $|a|^2$ and $|b|^2$

We first analyze the dynamical behavior before tunneling in the interval $S_1$: $0 \le \tau_1 < T$. To this end, we employ the change of variables and rescaling of the parameters similar to described in [19]. The change of variables

$$\Psi = |b|e^{i\Delta} = |\sin\theta|e^{i\Delta}, \ \Psi^* = |b|e^{-i\Delta} = |\sin\theta|e^{-i\Delta}, \ |\Psi| = |b|, \tag{4.3}$$

reduces system (4.2) to a single complex-valued equation

$$\frac{d\Psi}{d\tau_1} = 4i[\frac{1-\frac{1}{2}(\Psi^2+2|\Psi|^2)}{\sqrt{1-|\Psi|^2}} + 2\omega(\tau_2)\Psi - 8k\Psi|\Psi|^2], \ \Psi(0) = 0, \tag{4.4}$$

where $\omega(\tau_2) = 2k - g(\tau_2) = \omega_0 - g_1\tau_2$, $\omega_0 = 2k - g_0$. It is important to note that the frequency $\omega(\tau_2)$ directly depends on the coefficient $k$, thereby reflecting the effect of nonlinearity even in the linear approximation of Eq. (4.4).

Using the condition $|b| = |\Psi| << 1$ in $S_1$ (Fig. 6$b$), we define $\psi = \varepsilon^{-1/2}\Psi$ and introduce a proper time scale $s = \varepsilon^{-1/2}\tau_1$. Then, direct estimation shows that the coefficients of Eq. (4.4) allow rescaling $8k = \varepsilon^{-1/2}\kappa$, $2\omega_0 = \varepsilon^{-1/2}w_0$, where the parameters $\kappa$ and $w_0$ are of $O(1)$. Finally, we denote $\varepsilon^{3/2}g_1 = \beta^2/4$ keeping in mind that $\beta << 1$. When we substitute the variable $\psi$ and the rescaled coefficients into Eq. (4.4) and ignore the terms of order higher than $\varepsilon$, we obtain the equation

$$\frac{d\psi}{ds} = 4i[w(s)\psi + 1 - ½\varepsilon(\psi^2 + |\psi|^2) - \kappa\varepsilon\psi|\psi|^2],\ \psi(0) = 0, \tag{4.5}$$

where $w(s) = w_0 - \beta^2 s/2$. Thus we get a quasilinear equation and the earlier developed iteration procedure [11] can be employed to construct an approximate solution. As shown in [11], the initial iteration $\psi_0(s)$ is chosen as a solution of the linear equation

$$\frac{d\psi_0}{ds} = 4i[w(s)\psi_0 + 1],\ \psi_0(0) = 0, \tag{4.6}$$

$$\psi_0(s) = 4ie^{i\phi(s)}\Phi(s),\ \Phi(s) = \int_0^s e^{-i\phi(z)}dz\ ,\ \phi(s) = w_0 s - (\beta s)^2.$$

Simple algebra shows that

$$\psi_0(s) = 4i\beta^{-1}e^{i(\phi(s)-\alpha^2)}F(s), \tag{4.7}$$

where $h(s) = \beta s - \alpha$, $\alpha = w_0/\beta$, and

$$F(s) = \int_{-\alpha}^{h(s)} e^{ih^2}dh = [C(h(s)) + C(\alpha)] + i[S(h(s)) + S(\alpha)].$$

Here $F(s)$ is the complex-valued Fresnel integral, $C(h)$ and $S(h)$ are the cos- and sin-Fresnel integrals, respectively. If $\psi_0$ is found, the main approximation to the function $|b|$ is calculated as $|b_0| = \varepsilon^{1/2}|\psi_0|$. Note that the parameter $w_0$ depends on the coefficient of nonlinearity $k$, and thus, the behavior of the solution $\psi_0(s)$ is conditioned by the value of $k$.

The procedure analogous to that of [11] suggests that the first iteration $\psi_1$ should be found from the "linearized" equation

$$\frac{d\psi_1}{ds} = 4i\{[w(s) - \varepsilon\kappa|\psi_0(s)|^2]\psi_1 - \varepsilon(½\psi_0^2(s) + |\psi_0(s)|^2) + 1\},\ \psi_1(0) = 0, \tag{4.8}$$

which yields the solution

$$\psi_1(s) = \psi_0(s) + 4ie^{i\varphi(s)}\Phi_1(s), \tag{4.9}$$

$$\varphi(s) = \phi(s) - \varepsilon\kappa \int_0^s |\psi_0(z)|^2 \, dz \,,$$

$$\Phi_1(s) = \int_0^s e^{-i\varphi(z)} dz - \varepsilon \int_0^s e^{-i\varphi(z)} R(z) dz, \;\; R(z) = ½\psi_0^2(z) + |\psi_0(z)|^2.$$

Once the solution $\psi_1$ is derived, the first iteration to the function $|b|$ is calculated as $|b_1| = \varepsilon^{1/2}|\psi_1|$. The exact solution $|b(\tau_1)|^2$ and the iterations $|b_0(\tau_1)|^2$ and $|b_1(\tau_1)|^2$ in the interval $S_1$ are presented in Fig. 7*a*. Figure 7*a* clearly indicates that during the first half of the time-interval $S_1$ maximum difference between the linear approximation and the exact solutions is less than 15%. An increase of divergence at the second part of the interval $S_1$ is due to the different behavior of the exact and approximate solutions near the point of transition: while the exact solution undergoes a sudden increase at an instant of tunneling, the linear approximation tends to a certain limiting value.

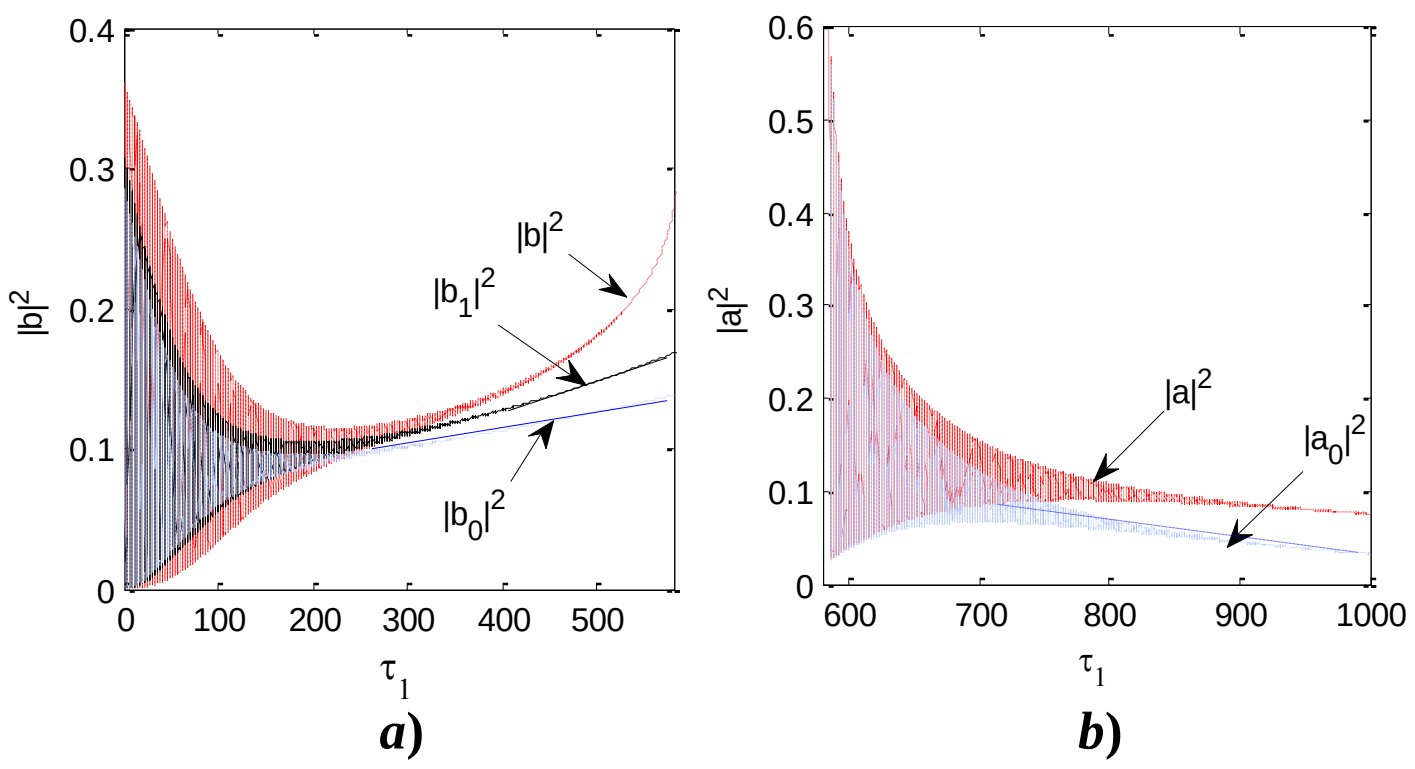

FIG. 7. Exact and approximate solutions before and after tunneling; (*a*): plots of the exact solution $|b|^2$ (red) and the iterations $|b_0|^2$ (light blue) and $|b_1|^2$ (black) in the interval $S_1$; (*b*): plots of the exact solution $|a|^2$ (red) and the iterations $|a_0|^2$ (light blue) in the interval $S_2$

The dynamical behavior in the interval $S_2$: $\tau_1 > T$ is studied in a similar way. Given $|a| << 1$, the variables analogous to (4.3) are introduced

$$Z = -\,|a|e^{i\Delta} = -\,|\cos\theta|e^{i\Delta}\,,\; Z^* = -\,|a|e^{-i\Delta} = -\,|\cos\theta|e^{-i\Delta}\,,\; |Z| = |a| \tag{4.10}$$

Transformations of the same sort that led to Eqs. (4.4) yield the following equation for $Z$:

$$\frac{dZ}{d\tau_1} = 4i[2\omega_1(\tau_2)Z - 1 + ½(Z^2 + |Z|^2) - 8\kappa Z|Z|^2],\; Z(T) = p_0, \tag{4.11}$$

with $\omega_1(\tau_2) = 2k + g(\tau_2) = \omega_{10} + g_1\tau_2$, $\omega_{10} = 2k + g_0$. As shown in Sec. II, the initial condition for Eq. (4.11) at $\tau_1 = T$ should be defined from the condition of coalescence of the stable and unstable points. This means that the initial value $\theta(T) = \theta_T$, which is calculated by Eq. (2.11), determines the initial values $|a(T)|^2 = \cos^2\theta_T$ and thus, $p_0 = |\cos\theta_T|$.

Given $|Z| << 1$ in $S_2$, we introduce the transformation $Z = \varepsilon^{1/2}z$, $\tau_1 = \varepsilon^{1/2}s$, $8k = \varepsilon^{-1/2}\kappa$, $2\omega_{10} = \varepsilon^{-1/2}w_{10}$ and denote $\varepsilon^{3/2}g_1 = \beta^2/4$, $\varepsilon^{-1/2}T = s_0$, $\varepsilon^{1/2}p_0 = p_{10}$. Substituting the rescaled quantities into Eq. (4.11) and ignoring the higher-order terms, we obtain the quasilinear equation similar to (4.5)

$$\frac{dz}{ds} = 4i[w_1(s)z - 1 + ½\varepsilon(z^2 + 2|z|^2) - \varepsilon\kappa z|z|^2],\ z(s_0) = p_{10}, \tag{4.12}$$

where $w_1(s) = w_{10} + \beta^2 s/2$ is an adiabatically increasing parameter. The initial iteration $z_0(s)$ is sought as a solution of the linear equation

$$\frac{dz_0}{ds} = 4i[w_1(s)z_0 - 1],\ z_0(s_0) = p_{10}, \tag{4.13}$$

$$z_0(s) = (p_1 - 4i\Theta(s))e^{i\delta(s)},\quad \Theta(s) = \int_0^s e^{-i\delta(r)}dr\,.$$

where $\delta(r) = 4[w_{10}r + (\beta r)^2]$. Using the representation $\delta(r) = h_1^2(r) - \alpha_1^2$, $h_1(r) = 2\beta r + \alpha_1$, $\alpha_1 = w_{10}/\beta$, we rewrite $\Theta(s)$ as

$$\Theta(s) = e^{i\alpha_1^2}F_1(s)/2\beta, \tag{4.14}$$

$$F_1(s) = \int_{\alpha_1}^{h_1(s)} e^{-ih^2}dh = [C(h_1(s)) - C(\alpha_1)] - i[S(h_1(s)) - S(\alpha_1)],$$

where $F_1(s)$ is the complex-valued Fresnel integral. Applying the inverse rescaling, we get the initial iteration to $|a|$ as $|a_0| = \varepsilon^{1/2}|z_0|$. As seen in Fig. 7*b*, the plot of $|a_0(\tau_1)|^2$ closely coincides with $|a(\tau_1)|^2$.

Now we calculate the stationary state $|\bar{a}_0| = \lim|a_0(\tau_1)|$ as $\tau_1 \to \infty$. In the limit of $s \to \infty$, $h_1(s) \to \infty$, the cos- and sin- Fresnel integrals are approximated as [27]

$$C(h_1(s)) \to \tfrac{1}{2}\sqrt{\tfrac{\pi}{2}},\ S(h_1(s)) \to \tfrac{1}{2}\sqrt{\tfrac{\pi}{2}}\,. \tag{4.15}$$

Since $\alpha_1 = w_{10}/\beta >> 1$ for $\beta << 1$, the terms $C(\alpha_1)$ and $S(\alpha_1)$ in (4.14) allows the approximation

$$C(\alpha_1) \approx \tfrac{1}{2}\left(\sqrt{\tfrac{\pi}{2}} + \frac{\sin\alpha_1^2}{\alpha_1}\right) + O\left(\frac{1}{\alpha_1^2}\right),\ S(\alpha_1) \approx \tfrac{1}{2}\left(\sqrt{\tfrac{\pi}{2}} - \frac{\cos\alpha_1^2}{\alpha_1}\right) + O\left(\frac{1}{\alpha_1^2}\right). \tag{4.16}$$

Combining (4.14) – (4.16), we find $F_1(s) \to i\, e^{-i\alpha_1^2}/2\alpha_1$, $\Theta(s) \to i/4w_{10}$, $z_0(s) \to (p_{10} + 1/w_{10})\, e^{i\delta(s)}$, and, therefore, $|z_0(s)| \to \bar{z}_0 = |p_1 + 1/w_{10}|$, as $s \to \infty$. Finally, we obtain

$$|\bar{a}_0| = \varepsilon^{1/2}\bar{z}_0 = |p_0 + 1/\omega_{10}|. \tag{4.17}$$

The resulting non-zero limiting value of $|\bar{a}_0|$ implies the existence of the residual energy depending on the initial condition $p_0$. Given $k = 0.65$, $g_0 = -0.5$, we obtain $|a_0| = 0.445$, $|a_0|^2 = 0.198$. It follows from (4.17) that in the interval $S_2$: $\tau_1 > T$ the energy of the coupled oscillator $e_2$ tends to the mean value $\bar{e}_2 = |\bar{b}_0|^2/2 = 1 - |\bar{a}_0|^2/2 = 0.901$. The quantity $\bar{e}_2$ may be interpreted as the amount of energy transferred from the initially excited oscillator to the coupled oscillator being initially at rest.

## V. TUNNELING IN STRONGLY NONLINEAR SYSTEMS

### A. Adiabatic transitions

As remarked previously, the dynamical behavior of system (4.2) directly depends on the likely transformation of the separatrix into the LPT at some critical values of $k$ and $g$. In this section, we examine strongly nonlinear systems, wherein the above-mentioned transformation is observed. The analysis of strongly nonlinear systems with constant detuning (Sec. II) demonstrates that, in terms of classical mechanics, the change of detuning in the “equivalent pendulum” (2.9) entails a passage from small oscillations to rotations (Figs. 2–4); physically, this corresponds to the transition from weak to strong energy exchange. In this section we extend this conclusion to strongly nonlinear systems with slowly-varying detuning. Furthermore, we expose an additional transition from rotations to large oscillations also associated with strong energy exchange.

Figure 8 demonstrates the evolution of the coordinates $\theta$ and $\Delta$ for the “equivalent pendulum” (4.2) with the parameters $k = 0.9$, $g_0 = -0.25$; $\varepsilon g_1 = 0.001$. Figure 8*a* depicts the time evolution of the phase $\Delta(t)$. The averaged value of $\Delta(t)$ is close to certain constants in the intervals $S_1$: $0 \le \tau_1 < T$ and $S_3$: $\tau_1 > T^*$ and linearly depends on the time in the interval $S_2$: $T \le \tau_1 < T^*$; the behavior of $\Delta(t)$ suggests the oscillatory motion in $S_1$ and $S_3$ and the rotational motion in $S_2$.

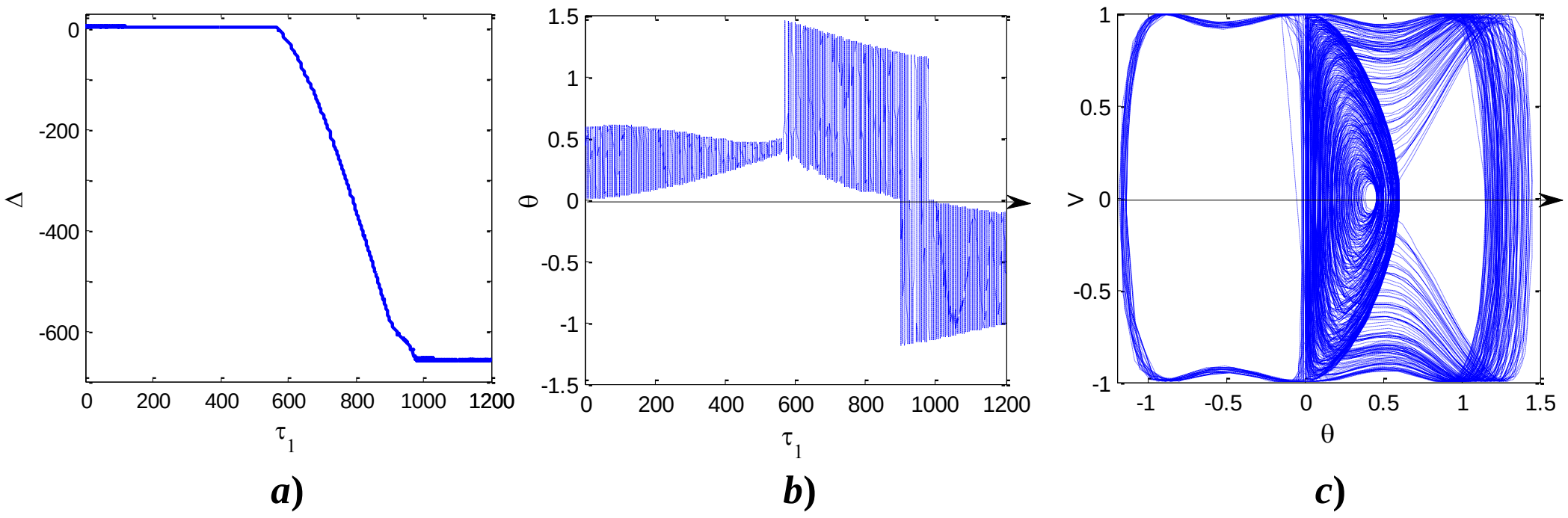

FIG. 8. Transient processes on the interval $0 \le \tau_1 \le 1200$; (*a*): phase $\Delta(\tau_1)$; (*b*): evolution of $\theta(\tau_1)$; (*c*): phase portrait in the plane $(\theta, V)$; the left half-plane corresponds to the interval $S_3$

The dynamical behavior of $\theta(\tau_1)$ and the phase portrait in the plane $(\theta, V = d\theta/d\tau_1)$ are shown in Fig. 8*b* and Fig. 8*c*, respectively. The phenomenon of energy localization is observed on the initial interval of motion $S_1$: $0 \le \tau_1 < T$; after tunneling, on the interval $S_2$: $T \le \tau_1 < T^*$, energy localization changes to intense exchange, associated with the transition from oscillations small amplitude to rotations; on the interval $S_3$: $\tau_1 > T^*$ a new transition from rotations to large oscillations is observed (the left half-plane in Fig. 8*c*). Note that tunneling occurs at $T = 580$ that corresponds to $g = 0.33$; this value coincides with the result obtained for the stationary system (Fig. 2) as well as with result of calculation by formula (2.11).

The adiabatic convergence to the transition point at the first stage of motion can be examined using the techniques of Sec. IV. In the current section, we analyze the dynamical behavior at the second stage $S_2$: $T < \tau_1 < T^*$, where the variation of $\theta(\tau_1)$ is large enough, and the asymptotic approach of Sec. IV is inapplicable.

We recall that the local minima and maxima of $\theta$ lie on the slowly varying envelops $Q^-(\tau_2)$ and $Q^+(\tau_2)$, respectively (Fig. 9). The stroboscopic method (see Appendix) leads to the following nonlinear equations for the envelopes:

$$2(\cos 2Q^- + k\sin 4Q^- - g(\tau_2)\sin 2Q^-)\frac{dQ^-}{d\tau_1} = \varepsilon g_1(\cos 2Q^+ - \cos 2Q^-), \tag{5.1}$$

$$2(-\cos 2Q^+ + k\sin 4Q^+ - g(\tau_2)\sin 2Q^+)\frac{dQ^+}{d\tau_1} = \varepsilon g_1(\cos 2Q^- - \cos 2Q^+),$$

with initial conditions $Q^- = \theta_0^-$, $Q^+ = \theta_0^+$ at $\tau_1 = T$. The point $\theta_0^-$ is defined as a point of coalescence of the stable and unstable states at the moment of tunneling; the point $\theta_0^+$ may be calculated by Eq. (5.5) in which $G_n = G_0 = g_0 + \varepsilon g_1 T$. If we take into account that the quantities $|Q^{-}|$ and $|\pi/2 - Q^{+}|$ are small enough (Fig. 9), then the initial approximation can be chosen as $Q_0^{-} = 0$, $Q_0^{+} = \pi/2$. Substituting $Q_0^{-}$, $Q_0^{+}$ into Eq. (5.10) and taking into account the initial conditions, we easily obtain the first approximation

$$Q_1^{\pm}(\tau_1) = \theta_0^{\pm} - \varepsilon g_1(\tau_1 - T). \tag{5.2}$$

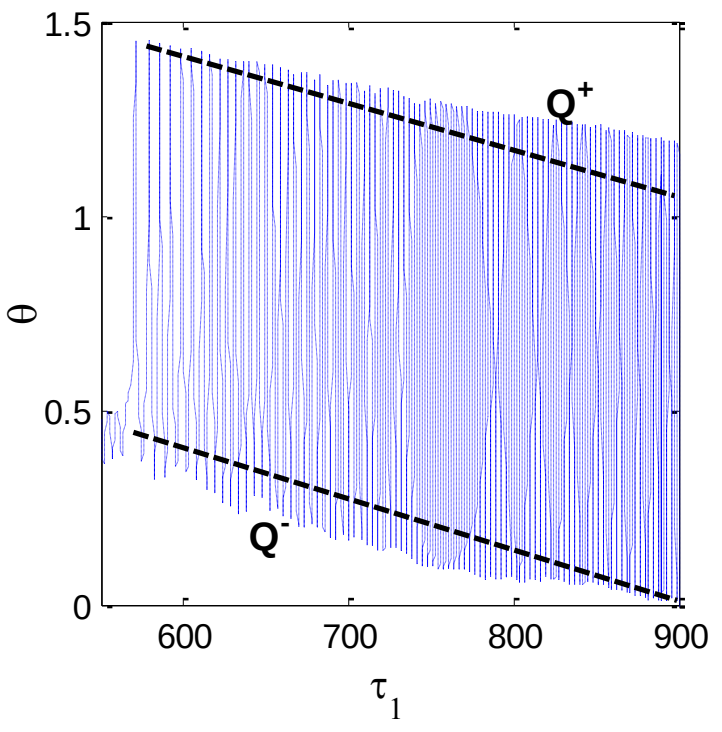

FIG. 9. Evolution of $\theta(\tau_1)$ on the interval $S_2$; black dashed lines correspond to approximate envelopes (5.11)

Figure 9 proves a good agreement between the approximate solution (5.2) and the precise (numerical) solution.

### B. Rapid transitions

As remarked previously, adiabatic tunneling characterized by an instant energy transfer from the excited to the coupled oscillator, requires a very large time for reaching the transition point from the initial state. Here we demonstrate the scenario of rapid tunneling, which may occur in the presence of the homoclinic separatrix coinciding with the LPT. We consider rapid tunneling as a passage through the separatrix and construct an explicit approximate solution describing this process.

Figure 10 presents the phase plots of the adiabatic systems with the coefficient of nonlinearity $k = 0.9$ and slowly-varying detuning. The bold lines in depict the homoclinic separatrix coinciding with the LPT in the conservative system with critical detuning $g^{*} = 0.1$.

We recall that by definition the initial conditions are given at the LPT at $\theta = 0$. As seen in Figs. 10*c* and 10*d*, the transition of the trajectory starting at $\theta = 0$ to the upper lever near $\theta = \pi/2$ is due to passage through the separatrix in a vicinity of the saddle point; since the process starts on the separatrix, the time to reach the transition point is small enough. This implies that the interval of motion $S_1$ before tunneling is short enough to neglect the change of detuning and ignore deviations of the starting point in each cycle of oscillations from $\theta = 0$. In turn, on the interval $S_2$, deviations of the starting point from $\theta = \pi/2$ become visible for sufficiently large $\tau_1$ but still remain negligible in the first approximation. This enables the interpretation of each cycle as motion along the LPT corresponding to a relevant value of detuning.

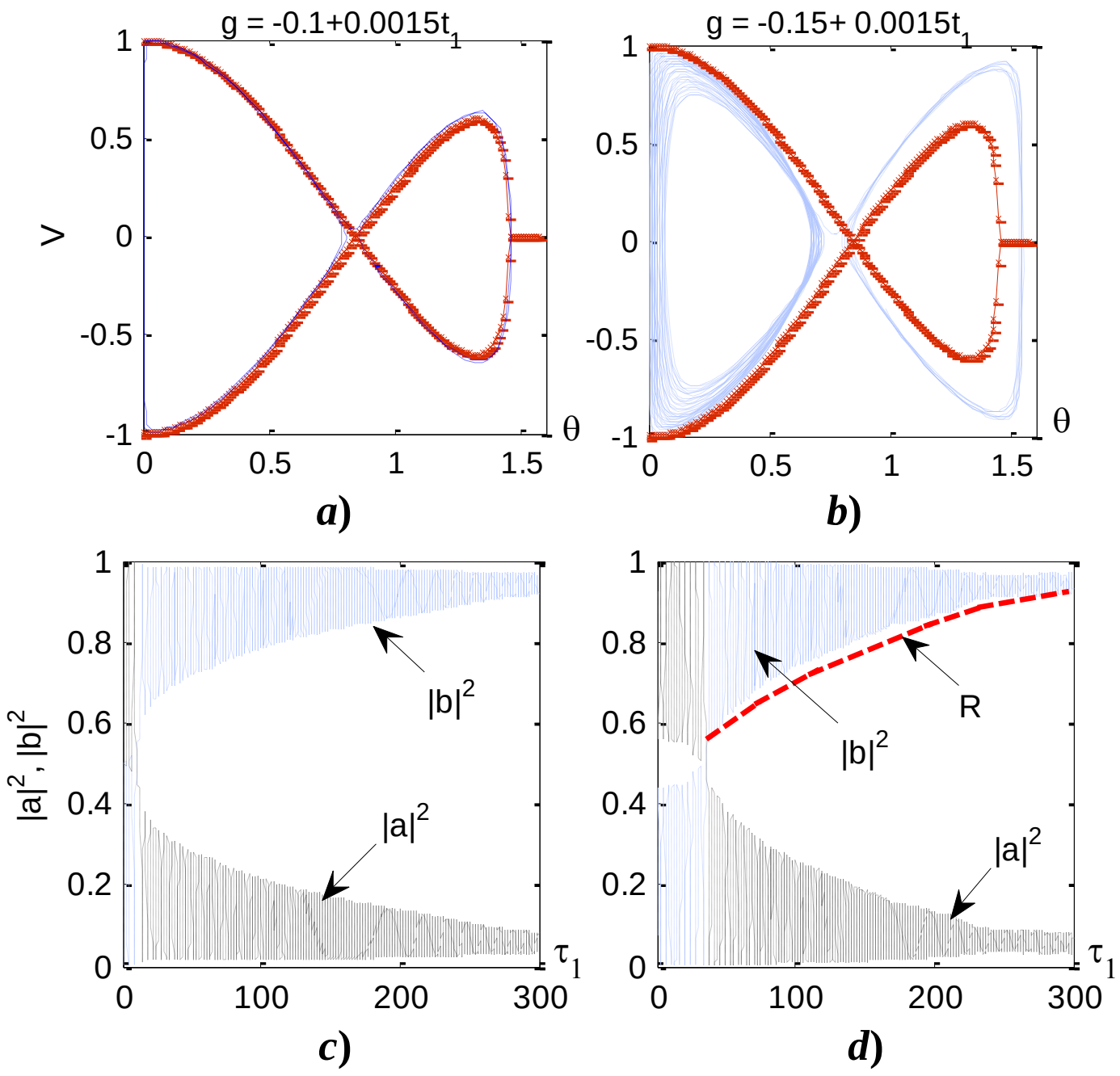

FIG. 10. Rapid tunneling for the system with $k$ = 0.9 and detuning $g(\tau_1)$ = –0.1 + 0.0015$\tau_1$ (*a*) and $g(\tau_1)$ = –0.15 + 0.0015$\tau_1$ (*b*). Bold solid lines in the top figures depict the separatrix of the system with constant critical detuning $g^*$ = – 0.1; bold dashed line R on Fig. *d* depicts the approximate envelope of $|b(\tau_1)|^2$ calculated by (5.13)

In order to construct a slowly-varying envelope of $\theta(\tau_1)$, we evaluate the extrema $\theta^*$of $\theta(\tau_1)$ on each cycle of motion. It follows from (2.9) that $V(\theta) = \pm\,[1 - (k\sin2\theta - g\tan\theta)^2]^{1/2}$ on the lower branch of the LPT with initial condition $\theta(0) = 0$ and $V(\theta) = \pm\,[1 - (k\sin2\theta + g\cot\theta)^2]^{1/2}$ on the upper branch with the initial condition $\theta(0) = \pi/2$. The extrema $\theta^*$ of the function $\theta(\tau_1)$ should be found from the condition $V(\theta) = 0$, yielding

$$k\sin2\theta - g\tan\theta = 1 \text{ on the lower branch of LPT,} \tag{5.12}$$

$$k\sin2\theta + g\cot\theta = 1 \text{ on the upper branch of LPT}$$

Since $\theta^* = \pi/4 + q$, $|q| <<1$, Eqs. (5.12) can be transformed into the linear equation

$$k - g(1 + 2q) = 1,\ q = ½[(k - 1)/g - 1] \text{ on the lower branch of LPT,} \tag{5.13}$$

$$k + g(1 - 2q) = 1,\ q = ½[1 - (1- k)/g] \text{ on the upper branch of LPT.}$$

Insertion of the time-dependent detuning $g(\tau_2)$ into Eqs. (5.13) provides adiabatic approximations of the slowly-varying envelopes. The envelope of the upper branch of the process $|a|^2 = \cos^2\theta$ for the system with the parameters $k = 0.9$, $g_0 = –0.15$, $\varepsilon g_1 = 0.0015$ is plotted in Fig. 10*b*. It is obvious that approximation (5.13) (bold dashed line) is close to the lower envelope of the process. The upper envelope is approximated by $\theta = \pi/2$, $|a| = 1$.

## V. CONCLUSIONS

In the paper we studied energy transfer in a system of two weakly coupled nonlinear oscillators in which the first oscillator with constant parameters is excited by an initial impulse, whereas the coupled oscillator with a time-dependent frequency is initially at rest. It was emphasized that the equations for the slowly varying envelopes in passage through resonance are identical to the equations of nonlinear Landau-Zener tunneling. This equivalence allows a unified approach to the study of such physically different processes as energy transfer in classical oscillatory systems under slow driving and nonlinear quantum Landau-Zener tunneling.

Three types of the nonlinear dynamical behavior are distinguished, namely, quasilinear, moderately nonlinearity, and strongly nonlinear one. It was shown that the previously exposed nonlinear analog of Landau–Zener tunneling between the energy levels is typical for moderately nonlinear systems, whereas strongly nonlinear systems exhibit a transition from the energy localization to a newly revealed cascade of the energy transfers. We found the relationships between the parameters determining a type of the dynamical behavior and then suggested a set of approximations to derive an analytical description of different types of the transient processes. Correctness of the constructed approximations is confirmed by numerical simulations.

## ACKNOWLEDGEMENTS

This work was supported by the Russian Foundation for Basic Research through the RFBR Grant # 10-01-00698. Also, the authors would like to thank anonymous reviewers for their comments and advice.

## Appendix

Equations (5.1) for the slowly varying envelopes are derived in this appendix. Let $T_n$ denotes a time-instant at which the function $\theta(\tau_1)$ achieves its *n-th* extremum $\theta_n$; a corresponding value of the phase $\Delta(\tau_1)$ is given by $\Delta_n = \pi n$. Using the stroboscopic method (see, e.g., [28]), we approximate a smooth function $g(\tau_2)$ by its step approximation

$$G(\tau_2) = \sum_{n=0}^{N} G_n[\eta(T_{n+1} - \tau_2) - \eta(T_n - \tau_2)], \tag{A.1}$$

where $G_n = g(T_n)$; $T_n$ denotes a time-instant at which the function $\theta(\tau_1)$ achieves its *n-th* extremum $\theta_n$; a corresponding value of the phase $\Delta(\tau_1)$ is given by $\Delta_n = \pi n$. It now follows from (A.1) that $G(\tau_2) = G_n$ in each interval $I_n$: $T_n \leq \tau_2 < T_{n+1}$. The step approximation allows us to replace equations (4.2) by the system with constant detuning in each interval $I_n$, namely,

$$\frac{d\theta}{d\tau_1} = \sin\Delta, \tag{A.2}$$

$$\frac{d\Delta}{d\tau_1} = 2(\cos\Delta + 2k\sin 2\theta)\cot 2\theta - 2G_n.$$

By analogy with (3.1), one can conclude that in the interval $I_n$ system (A.2) preserves the first integral of motion

$$K_n = (\cos\Delta + k\sin 2\theta)\sin 2\theta + G_n\cos 2\theta\,. \tag{A.3}$$

The parameter $K_n$ on both ends of the interval $I_n$, namely at $\theta_n = \theta(T_n)$, $\Delta_n = \pi n$ and $\theta_{n+1} = \theta(T_{n+1})$, $\Delta_{n+1} = \pi(n+1)$, is expressed as

$$K_n = [(-1)^n + k\sin 2\theta_n]\sin 2\theta_n + G_n\cos 2\theta_n = \Psi_n(\theta_n), \tag{A.4}$$

$$K_n = [(-1)^{n+1} + k\sin 2\theta_{n+1}]\sin 2\theta_{n+1} + G_n\cos 2\theta_{n+1} = \Phi_n(\theta_{n+1}),$$

where $\Psi_n(\theta_n) = \Phi_n(\theta_{n+1})$. Here and below the functions $\Psi_n$ and $\Phi_n$ are defined at the left and right ends of the interval $I_n$, respectively.

Similarly, in the interval $I_{n+1}$: $T_{n+1} \leq \tau_2 < T_{n+2}$ system (A.2) preserves the integral

$$K_{n+1} = (\cos\Delta + k\sin 2\theta)\sin 2\theta + G_{n+1}\cos 2\theta. \tag{A.5}$$

such that

$$K_{n+1} = [(-1)^{n+1} + k\sin 2\theta_{n+1}]\sin 2\theta_{n+1} + G_{n+1}\cos 2\theta_{n+1} = \Psi_{n+1}(\theta_{n+1}), \tag{A.6}$$

$$K_{n+1} = [(-1)^{n} + k\sin 2\theta_{n+2}]\sin 2\theta_{n+2} + G_{n+1}\cos 2\theta_{n+2} = \Phi_{n+1}(\theta_{n+2}),$$

with $\Psi_{n+1}(\theta_{n+1}) = \Phi_{n+1}(\theta_{n+2})$. Here $\theta_{n+1}$ and $\theta_{n+2}$ are the initial and end points of $I_{n+}1$, corresponding to $\Delta = \pi(n+1)$ and $\pi(n+2)$, respectively. The integrals $K_r$, $r = n + r$, $r > 1$, may be expressed in a similar way. It now follows from Eqs. (A.4) - (A.6) that

$$K_{n+1} - K_n = \Phi_{n+1}(\theta_{n+2}) - \Psi_n(\theta_n) = \Psi_{n+1}(\theta_{n+1}) - \Phi_n(\theta_{n+1}). \tag{A.7}$$

Let $\theta_{n+2} - \theta_n = \delta\theta_n$, $G_{n+1} - G_n = \delta G_n$. Substituting (A.4) – (A.6) into (A.7) and considering only the terms linear in $\delta\theta_n$, $\delta G_n$, we obtain the difference approximations

$$2((-1)^n\cos 2\theta_n + k\sin 4\theta_n - G_n\sin 2\theta_n)\delta\theta_n = (\cos 2\theta_{n+1} - \cos 2\theta_n)\delta G_n, \tag{A.8}$$

In the same way, a comparison of the integrals $K_{n+2}$ and $K_{n+1}$ in the intervals $I_{n+2}$ and $I_{n+1}$ yields the following difference equation:

$$2((-1)^{n+1}\cos 2\theta_{n+1} + k\sin 4\theta_{n+1} - G_{n+1}\sin 2\theta_{n+1})\delta\theta_{n+1} = (\cos 2\theta_{n+2} - \cos 2\theta_{n+1})\delta G_{n+1}. \tag{A.9}$$

For definiteness, we suppose for a given $n$ the points $\theta_n$ and $\theta_{n+1}$ correspond to a local minimum and a consequent maximum of $\theta(\tau_1)$, respectively, that is $n = 2k$. If we recall that the local minima and maxima of $\theta(\tau_1)$ lie on the slowly varying envelops $Q^-(\tau_2)$ and $Q^+(\tau_2)$, respectively, then in the limit of small differences Eqs. (A.8) and (A.9) can be replaced by the following system of differential equations for the slowly-varying envelops:

$$2(\cos 2Q^- + k\sin 4Q^- - g(\tau_2)\sin 2Q^-)\frac{dQ^-}{d\tau_1} = \varepsilon g_1(\cos 2Q^+ - \cos 2Q^-), \tag{A.10}$$

$$2(-\cos 2Q^+ + k\sin 4Q^+ - g(\tau_2)\sin 2Q^+)\frac{dQ^+}{d\tau_1} = \varepsilon g_1(\cos 2Q^- - \cos 2Q^+),$$

with initial conditions $Q^- = \theta_0^-$, $Q^+ = \theta_0^+$ at $\tau_1 = T$. The point $\theta_0^-$ is defined as a point of coalescence of the stable and unstable states at the moment of tunneling; the point $\theta_0^+$ may be calculated by Eq. (A.5) in which $G_n = G_0 = g_0 + \varepsilon g_1 T$.

## References

[1] G. Kopidakis, S. Aubry, and G.P. Tsironis, Phys. Rev. Lett. **87,** 165501 (2001).

[2] S. Aubry, G. Kopidakis, A.M. Morgante, and G.P. Tsironis, Physica B **296**, 222 (2001).

[3] P. Maniadis, G. Kopidakis, and S. Aubry, Physica D **188,** 153 (2004).

[4] L. Vazquez, R. MacKay, and M.P. Zorzano (Eds.), *Localization and Energy Transfer in Nonlinear Systems* (World Scientific, Singapore, 2003).

[5] T. Dauxois, A. Litvak-Hinenzon, R. MacKay, A. Spanoudaki, (Eds.), *Energy Localization and Transfer*, (World Scientific, Singapore, 2004).

[6] A.F. Vakakis, O. Gendelman, L.A. Bergman, D.M. McFarland, G. Kerschen, and Y.S. Lee, *Passive Nonlinear Targeted Energy Transfer* in *Mechanical and Structural Systems* (Springer-Verlag, Berlin New York, 2008).

[7] L.I. Manevitch and O.V. Gendelman, *Tractable Models of Solid Mechanics: Formulation, Analysis and Interpretation* (Springer-Verlag, Berlin Heidelberg, 2011).

[8] L.I. Manevitch, Yu.A. Kosevich, M. Mane, G.M. Sigalov, L.A. Bergman, and A.F. Vakakis, Int. J. Non-Linear Mech. **46,** 247 (2011); doi:10.1016/j.ijnonlinmec.2010.08.010.

[9] Yu.A. Kosevich, L.I. Manevitch, and E.L. Manevitch, Physics-Uspekhi, **53**, 1281 (2010).

[10] A. Kovaleva, L. Manevitch, and Yu. Kosevich, Phys Rev E **83**, 026602 (2011).

[11] A. Kovaleva, L. Manevitch, Phys Rev E **85**, 016202 (2012).

[12] F. Trimborn, D. Witthaut, V. Kegel, and H. J. Korsch, New J. Phys. **12**, 05310 (2010).

[13] N. Sahakyan, H. Azizbekyan, H. Ishkhanyan, R. Sokhoyan, and A. Ishkhanyan, Laser Physics **20**, 291 (2010).

[14] R. Khomeriki, Eur. Phys. J. D **61**, 193-197 (2011).

[15] R. Schilling, M. Vogelsberger, and D. A. Garanin, J. Phys. A: Math. Gen. **39** 13727 (2006).

[16] O.Zobay, B.M. Garraway, Phys. Rev. A **61**, 033603 (2000);

[17] J. Liu, L. Fu, B-Y Ou, S-G Chen, D-Il Choi, B. Wu, and Q, Niu, Phys. Rev. A **66**, 023404 (2002).

[18] A.P. Itin, A.A. Vasiliev, G. Krishna, and S. Watanabe, Physica D **232**, 108 (2007).

[19] A.P. Itin and P.Törmä, arXiv:0901.4778v1 (2010).

[20] A. Ishkhanyan, M. Mackie, A. Carmichael, P.L. Gould, and J. Javanainen, Phys. Rev. A **69**, 043612 (2004).

[21] E. Pazy, I Tikhonenkov, Y. B. Band, M. Fleischhauer, and A. Vardi, Phys. Rev. Lett. **95**, 170403 (2005)

[22] L. D. Landau, Phys. Z. Sowjetunion **2**, 46 (1932).

[23] C. Zener, Proc. R. Soc. London, A **137**, 696 (1932).

[24] A. Smerzi, S. Fantoni, S. Giovanazzi, and S. R. Shenoy, Phys. Rev. Lett. **79**, 4950 (1997).

[25] S. Raghavan, A. Smerzi, S. Fantoni, and S. R. Shenoy, Phys. Rev. A **59**, 620 (1999).

[26] L. I. Manevitch, Arch. Appl. Mech. 77, 301 (2007).

[27] I. S. Gradshteyn and I.M. Ryzhik, *Tables of Integrals, Series, and Products*, 6th ed. (Academic Press, San Diego, CA, 2000).

[28] N. Minorsky, *Nonlinear Oscillations* (Robert E. Kriger Publ. Comp., Huntington NY, 1974).